\documentclass[11pt,a4paper]{article}
\usepackage{amsmath,amssymb}
\usepackage{epsfig,graphicx}

\begin{document}

\title{\bf Study of \boldmath $K^-$$ \rightarrow $$\pi^{0} e^{-} 
\overline\nu_{e}$$ \gamma $ decay  with ISTRA+ setup.}

\author{S.A. Akimenko$^b$, 
V.~N.~Bolotov$^a$,
G.~I.~Brivich $^b$, 
K.~V.~Datsko$^b$, 
V.~A.~Duk$^{a}$, \\
A.~P.~Filin$^b$,
E.~N.~Guschin$^a$,
A.~V.~Inyakin$^b$,
V.~F.~Konstantinov$^b$,\\
A.~S.~Konstantinov$^b$,
I.~Y.~Korolkov$^b$, 
S.~V.~Laptev$^{a}$, 
V.~A.~Lebedev$^{a}$,\\
V.~M.~Leontiev$^b$, 
A.~E.~Mazurov$^{a}$,
V.~P.~Novikov$^b$, 
V.~F.~Obraztsov$^b$,\\
V.~A.~Polyakov$^b$, 
A.~Yu.~Polyarush$^{a}$, 
V.~E.~Postoev$^{a}$,
V.~I.~Romanovsky$^b$,\\ 
V.~I.~Shelikhov$^b$,
O.~G.~Tchikilev$^b$,
V.~A.~Uvarov$^b$,
O.`P.~Yushchenko$^b$
\\ 
$^a$ \small{\em Institute for Nuclear Research} \\
\small{\em INR RAS, prospekt 60-letiya Oktyabrya 7a, Moscow 117312, Russia } \\
$^b$ \small{\em Institute for High Energy Physics} \\
\small{Protvino, Russia}
}

\date{10.10.2006}
\maketitle

\vspace{1.0cm}

\begin{abstract}
Results of study of the $K^- \rightarrow \pi^{0} e
\overline\nu \gamma $ decay
at ISTRA+ setup are presented.
4476 events of this decay have been observed.
The branching ratio (R) is found to be
$R= \frac{Br(K^- \rightarrow \pi^{0} e^{-} \overline\nu_{e}
\gamma)}
{Br(K^- \rightarrow \pi^{0} e^{-} \overline\nu_{e})}
  =(1.81\pm0.03(stat)\pm0.07(syst) )\cdot10^{-2}$ for $E^{*}_{\gamma}>10MeV$ 
and $\theta^{*}_{e\gamma} >
10^{\circ}$.
For comparison with previous experiment the branching ratio
with cuts
$E^{*}_{\gamma}>10$MeV,
$0.6<cos\theta^{*}_{e \gamma}<0.9$ is measured
 $ R= \frac{Br(K^- \rightarrow \pi^{0} e^{-} \overline\nu_{e} \gamma) }
{Br(K^- \rightarrow \pi^{0} e^{-} \overline\nu_{e})}
=(0.47\pm0.02(stat)\pm0.03(syst))\cdot10^{-2}.$
For the cuts  $E^{*}(\gamma) > 30 MeV$ and $\theta^{*}_{e\gamma} >
20^{\circ}$,
used in most theoretical papers
$ Br = (3.06\pm0.09\pm0.14)\cdot 10^{-4}$.
For the asymmetry $A_{\xi}$(for the same cuts as in Table.2) we get
$A_{\xi} = -0.015 \pm 0.021$.
At present time it is the best estimate of this asymmetry.
\end{abstract}

\%newpage

\section{ Introduction }  
\large
The decay 
$K^- \rightarrow \pi^{0} e^{-} \overline\nu \gamma $
provides fertile testing ground  for the Chiral Perturbation Theory (ChPT)
\cite{v1,v2}.
$K^- \rightarrow \pi^{0} e \overline\nu \gamma $ decay amplitudes
are calculated at order ChPT $O(p^{4})$ in \cite{v1}, and
branching ratios are evaluated in \cite{v3}.
Recently next-to-leading $O(p^{6})$ corrections were calculated for the 
corresponding neutral kaon decay \cite{v4}.

The matrix element for  
$K^- \rightarrow \pi^{0} e \overline\nu 
\gamma $ has general structure
\begin{eqnarray}
T =\frac{G_{F}}{\sqrt{2}}{e}V_{us}\varepsilon^{\mu} 
(q)\Biggl\{(V_{\mu\nu}
- A_{\mu\nu})\overline{u}(p_{\nu})\gamma^{\nu}(1 - \gamma_{5})v(p_{l}) 
\\
\nonumber
\qquad\qquad\qquad + \frac 
{F_{\nu}}{2p_{l}q}\overline{u}(p_{\nu})\gamma^{\nu}(1 - 
\gamma_{5})(m_{l}-\not{{p}_{l}}-\not{q})\gamma_{\mu}v(p_{l})\Biggr\}\equiv
{\epsilon^{\mu}} A_{\mu}.
\end{eqnarray}

First term of the matrix element  describes Bremsstrahlung
 of kaon and direct emission(Fig.1a). The 
lepton Bremsstrahlung is presented by second term in r.h.s. of  
Eq(1) and (Fig.1b).

\begin{figure}[!hb]
\centering
\vspace*{-1.5cm}
\centerline{\epsfig{file=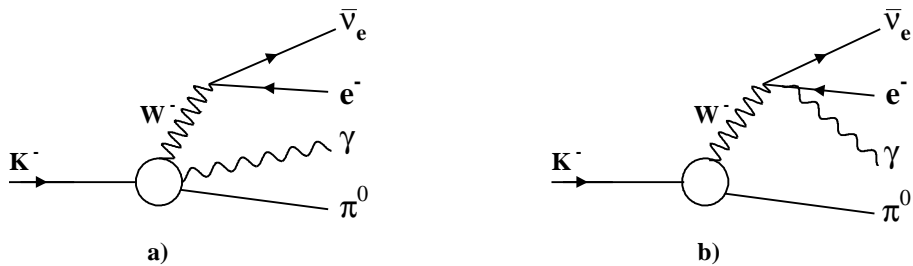,width=15cm }}
\vspace*{-1.5cm}
\caption{\em Diagrammatic representation of the
$K_{l3\gamma}$  amplitude.}
\label{figure:diag1}
\end{figure}

The $K^- \rightarrow \pi^{0} e \overline\nu \gamma $ decay
is one of kaon decays where new physics beyond the SM
can be probed. 
This decay is especially interesting as it is sensitive to T-odd
contributions. According to  CPT-theorem  observation  of T violation   
is equivalent to observation of CP-violating effects.
CP violation is a subject of continuing interest in K and B meson decays.

In the SM the source of CP violation is given by the phase 
in the CKM matrix\cite{v5,v6,v7}. 
However it has been argued that this source is not
enough to explain the observed baryon asymmetry of the universe
and new sources of CP violation have to be introduced\cite{v8}.

Important experimental observable used in CP-violation searches is
the T-odd correlation for 
$K^- \rightarrow \pi^{0} e 
\overline\nu \gamma $ decay defined as
\begin{equation}
\xi_{\pi e \gamma} = \frac{1} {M^{3}_{K}} 
p_{\gamma}\cdot[p_{\pi}\times
p_{e}]
\end{equation}
First suggestion to investigate T-odd triple-product correlations 
was done in\cite{v9}

To establish the presence of a nonzero triple-product correlations, one
constructs a T-odd asymmetry of the form

\begin{equation}
A_{\xi}= \frac {N_{+}-N_{-}} {N_{+}+N_{-}}
\end{equation}
Where $N_{+}$ and $N_{-}$ are number of events with
$\xi>0$ and $\xi<0$

T-odd correlation vanishes at tree level of SM\cite{v10}, but the SUSY 
theory gives 
rise 
to CP-odd(T-odd) observables already at tree level\cite{v11,v12,v13}.
T-odd asymmetry value for \\ 
$SU(2)_{L}$$ \times$$ SU(2)_{R}$$ \times$$ U(1)$ model 
and scalar  models was estimated in Ref\cite{v14}.

In this letter we present first results of the analysis of the 
$K^- \rightarrow \pi^{0} e
\overline\nu \gamma $ data accumulated by ISTRA+ experiment  during the 
2001 run.

\section{ ISTRA+ setup } 

\begin{figure}[!hb]
\centering
 \includegraphics[angle=90, width=15cm]{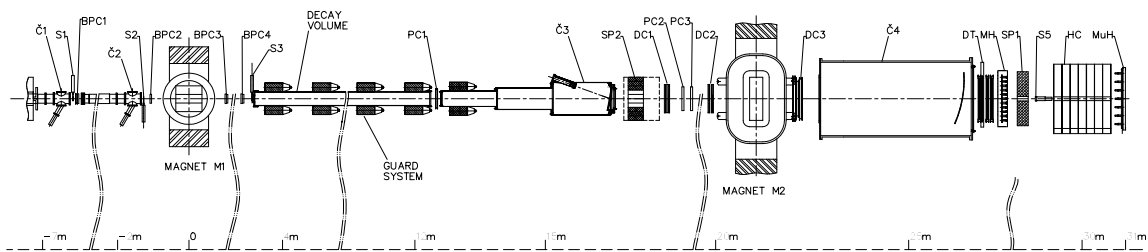}
  \caption{\em   The side view of the ISTRA+ detector}
  \label{figure:ust}
\end{figure}

The experiment was performed using ISTRA+ detector
which is modernized  ISTRA-M detector \cite{d1}.
ISTRA+ detector is located in a negative unseparated beam.
The measurement of the beam particles, deflected by the beam magnet M1
is performed by four beam proportional chambers $BPC_{1} \div BPC_{4}$.
The beam momentum is $\sim25 GeV$ with $\Delta p/p  \sim1.5\%$.
Admixture of $K^-$ in the beam is $\sim3\%$. The beam intensity is
$\sim3\cdot 10^6$ per 1.9 sec  U-70 spill.
The kaon identification is performed by  $\check{C_{0}}
\div \check{C_{2}}$ threshold $\check{C}$-counters ($\check{C_{0}}$ is not 
shown in Fig.2).

The decay products are deflected by the spectrometer magnet M2 with
the field integral of 1Tm. The track measurement is performed by 1-mm-step
proportional chambers ($PC_{1} \div PC_{3}$), 2-cm-cell drift chambers
($DC_{1}  \div DC_{3}$), and by four planes of the 2-cm-diameter drift tubes DT. 
The photons are  measured by lead-glass electromagnetic
calorimeter $SP_{1}$ which consists of 576 counters. The counter
transverse size is $5.2 \times 5.2$ cm and length is about 15 $X_{0}$.
To veto low energy photons the decay volume is surrounded
by eight lead-glass rings.
Lead-glass electromagnetic calorimeter $SP_{2}$ is also used as a part 
of the veto system.

\section{Event selection }

During physics run in November-December 2001 350M events were logged on 
tapes.
This information is complemented by 260M events generated with Geant3
\cite{e1}.
The Monte Carlo simulation includes a realistic description of the experimental
setup: the decay volume entrance windows, the track chamber windows,
gas mixtures, sense wires and cathode structures, Cherenkov counters mirrors and gas
mixtures, the showers development in the electromagnetic calorimeters, etc.
The detailed discussion of the simulation and  reconstruction procedure 
is given in our previous publications \cite{e2,e3}.

Events with one negative track detected in tracking system
and four showers detected in  electoromagnetic  calorimeter SP1
are selected as candidates for  $K^- \rightarrow \pi^{0} e \nu \gamma$
decay. One of this showers must be associated with the charged track.

Events with vertex inside interval $400 < z < 1650$ cm, and
transverse radius less than 10cm is selected for further analysis.

The probability of the vertex fit, $CL(\chi^{2})$, is required 
to be more than $10^{-4}$.
Absence of signals in veto system above noise threshold is required.

The electron identification is done using  E/P ratio of
the energy of the cluster associated with the track to momentum of this
track given by tracking system. This ratio must be inside interval 
0.80-1.15(see Fig.3). 
Another cut used for the suppression of the $\pi^{-}$ contamination  
is that    on the distance between the charged track extrapolation to the
front plane of the electoromagnetic detector and the nearest shower.
This distance must be less than 2,5 cm.

The effective mass $m( \gamma \gamma)$  within 
$\pm30$ MeV from $\pi^{0}$ table mass (Fig.4) is required.

At the end, the convergence of the 
2C $K^-$$ \rightarrow $$\pi^{0} e^{-}
\overline\nu_{e}$$ \gamma $ kinematic fit  is required.

\begin{figure}
\begin{minipage}[t]{0.45\textwidth}
\centering
\includegraphics[scale=.75]{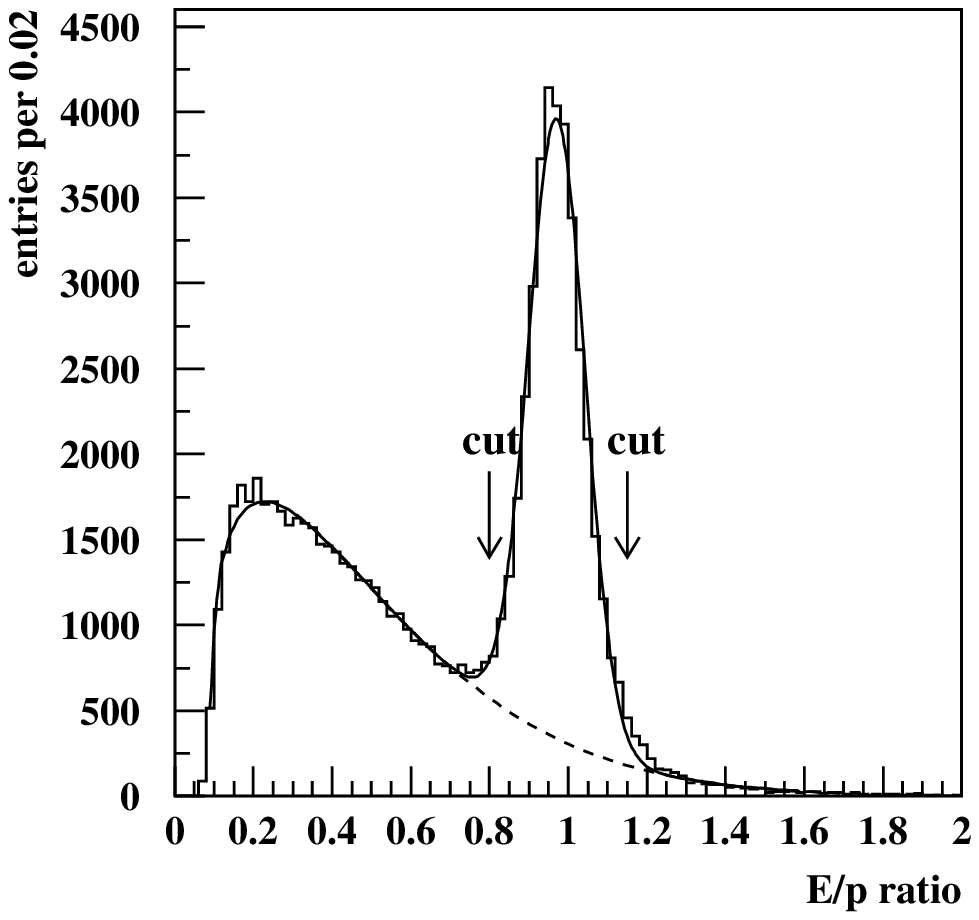}
  \caption{\em  
E/P ratio for the real data. Dotted line is our fit of background.}
  \label{figure:piz1}
\end{minipage}  
\hspace{1cm}
\begin{minipage}[t]{0.45\textwidth}
\centering
\includegraphics[scale=.75]{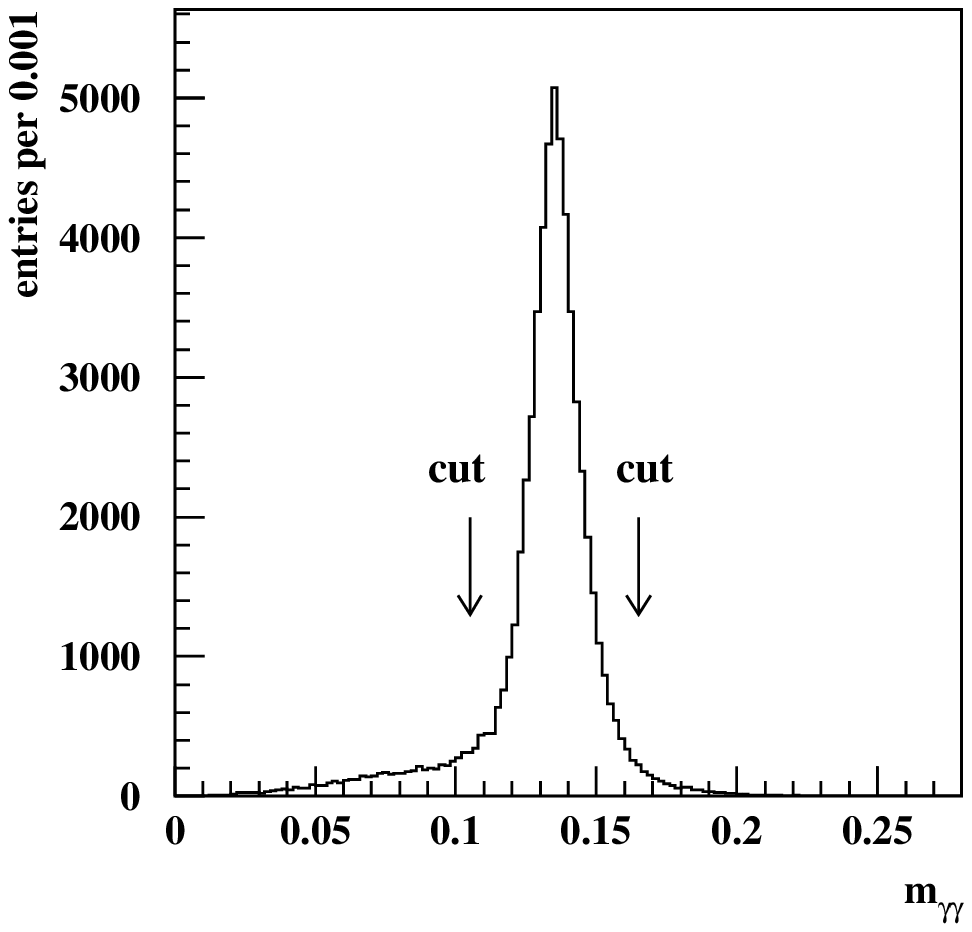}
  \caption{\em  $\gamma \gamma$ mass for the real data.}
  \label{figure:piz2}
\end{minipage}
\end{figure}

\section{Background suppression }

The main background decay channels 
for the decay $K^-$$ \rightarrow $$\pi^{0} e^{-}
\overline\nu_{e}$$ \gamma $
are:

 (1) $K^- \rightarrow \pi^{-} \pi^{0} \pi^{0} $  where one  
of the $\pi^{0}$ photons is not detected  and $\pi^{-}$
decays to $e \nu$   or is misidentified  as an electron.

(2) $K^- \rightarrow \pi^{-} \pi^{0}  $ with  ``fake  photon''   
and $\pi^{-}$ decayed or misidentified as electron. 
Fake photon clusters can come from $\pi$-hadron interaction in 
the detector,
external bremsstrahlung upstream of the magnet, accidentals.
All these sources are included in our MC calculations.

(3) $K^- \rightarrow \pi^{0} e \nu  $ with  extra photon.     
The main source of  extra photons is an electron interactions in 
the detector.

(4) $K^- \rightarrow \pi^{-} \pi^{0} \gamma$  when $\pi^{-}$  decays or
is mis-identified as an electron.

(5) $K^- \rightarrow \pi^{0} \pi^{0}  e \nu$  when one $\gamma$ is lost

From Fig.3 it is seen that  in raw data background 
contamination from channels with charged pion in final state is about 
15\%.

Requirement on the missing energy in the decay reduces 
mainly background channel(4).

Cut1: \hspace{0.8cm} $E_{miss} > 0.5 GeV$

For the suppression of the background channels (1-5)   we use a cut on the 
missing mass squared 

\hspace{1.0cm} $M^{2}$$(\pi^{0} e \gamma)$ = $(P_{K} - 
P_{\pi^{0}} - P_{e} - P_{\gamma})^2$. 

For the signal events this variable corresponds to the square of the neutrino 
mass
and must be zero within measurement accuracy (see Fig.5).

Cut2: \hspace{0.8cm} $-0.01 < M^{2}(\pi^{0} e^{-} \gamma) < 0.01$

\begin{figure}
\includegraphics[scale=.75]{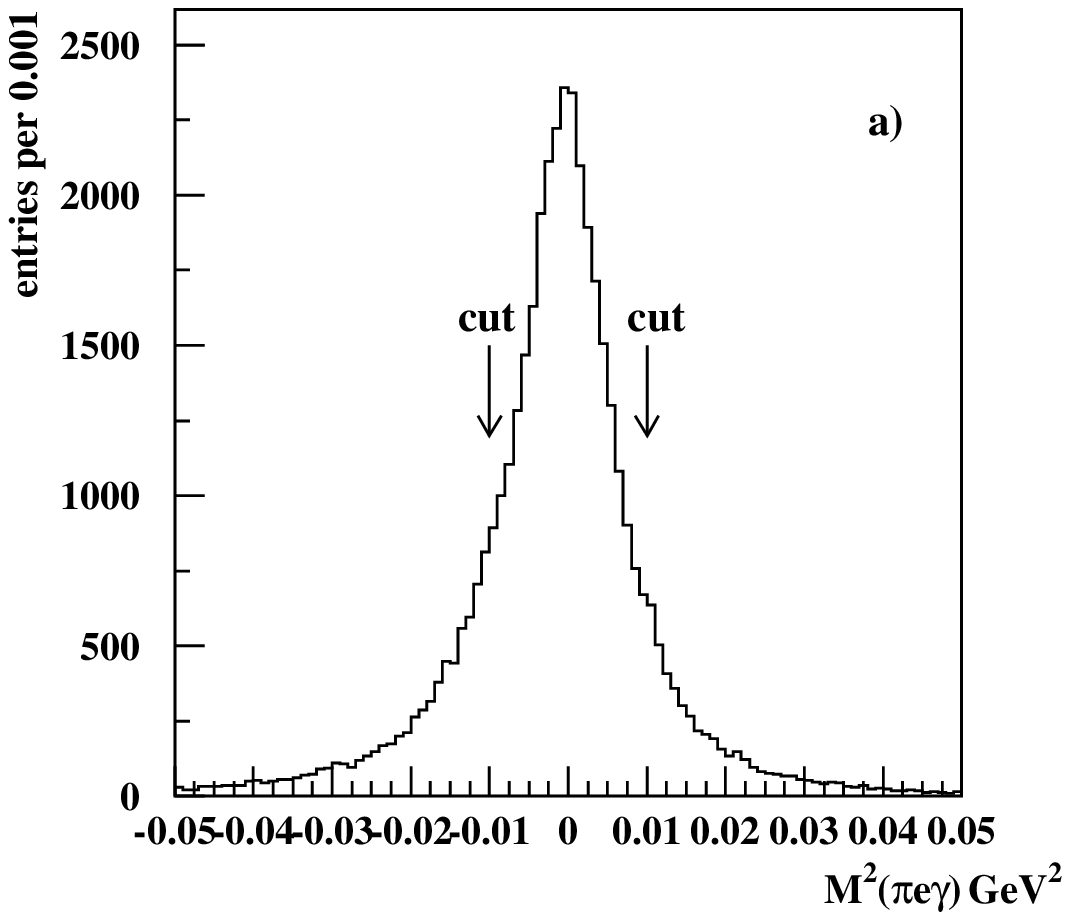}
\includegraphics[scale=.75]{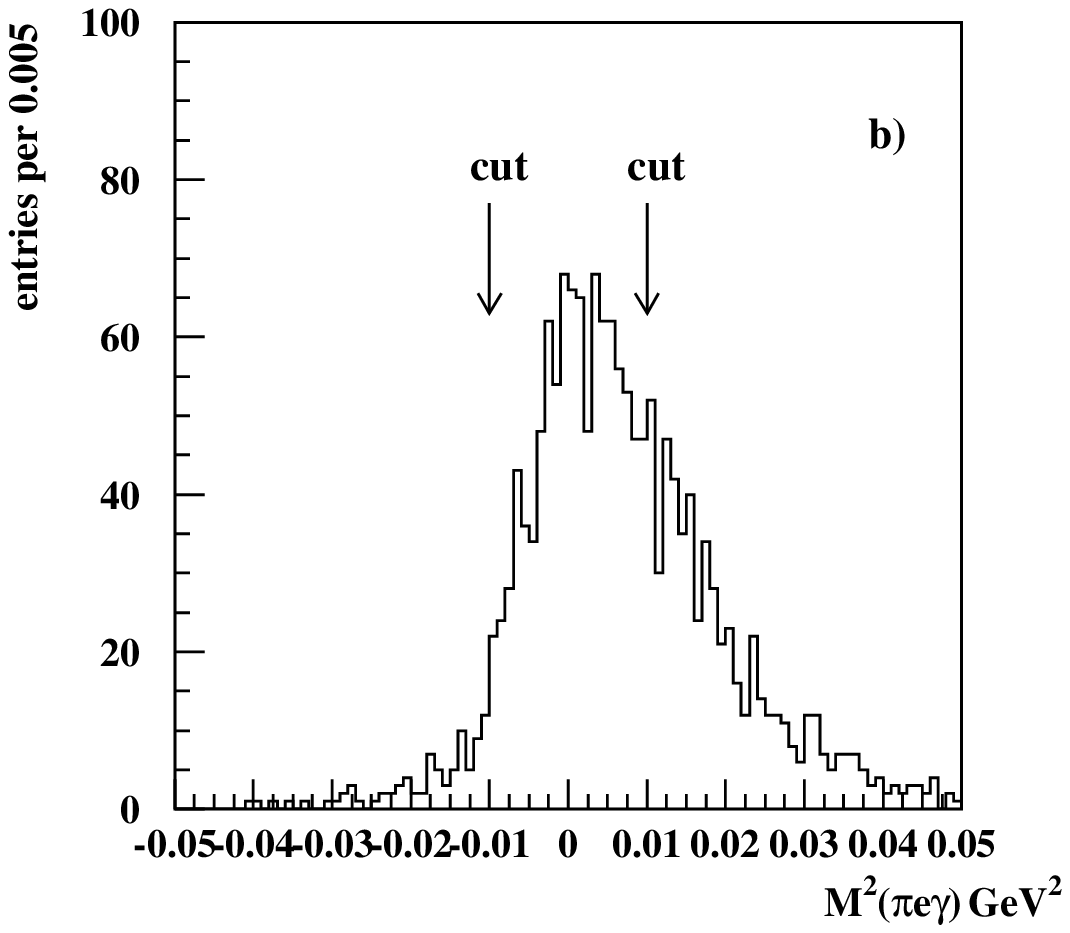}
  \caption{\em Missing mass $M^{2}(\pi^{0} e^{-} \gamma)$ distribution;
 a) for the real data b) for the background channel(1).}
  \label{figure:piz3}
\end{figure}


For the suppression of the background channel(1) we also use a  cut on the 
missing mass squared  
$M^{2}(\pi^{-} \pi^{0}) = (P_{K} - P_{\pi^{-}} - P_{\pi^{0}})^2$  

For the background(1)  events this variable corresponds to  
$\pi^{0}$ mass, 
for the signal  events distribution of this variable is rather wide (see 
Fig.6).

Cut3: \hspace{0.8cm}  The events with   
$0.009 < M^{2}(\pi^{-}\pi^{0}) < 0.024 $ are cutted out.

\begin{figure}
\includegraphics[scale=.72]{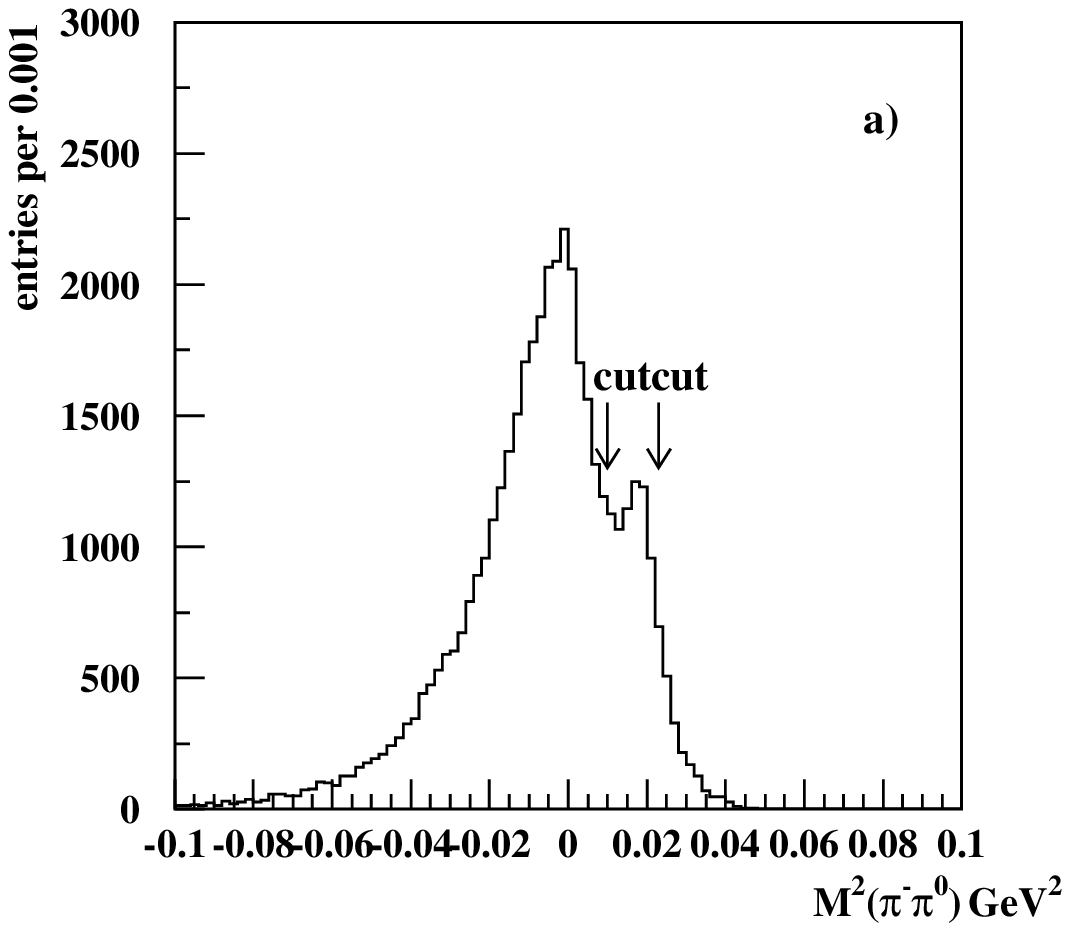}
\includegraphics[scale=.72]{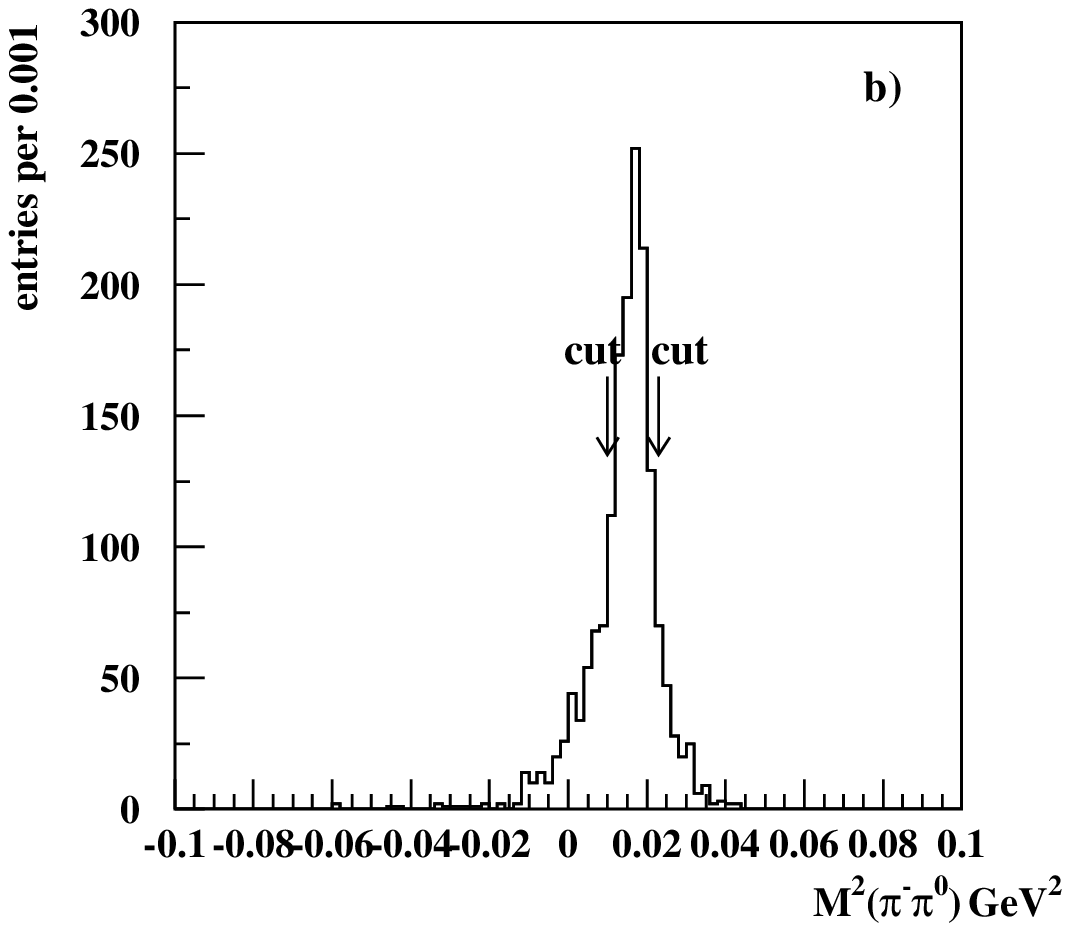}

  \caption{\em missing mass distribution
$M^{2}(\pi^{-}\pi^{0})$ a) for real data;
b) for the background (1)}
  \label{figure:piz5}
\end{figure}

The dominant background to $K_{e3\gamma}$ arises from 
$K_{e3}$ with extra photon. 
The background (3) is suppressed by requirement on the
 angle between electron and photon
in the laboratory frame  $\theta_{e \gamma}$  (see
Fig.7)
The distribution of the $K_{e3}$-background events
has very sharp peak at zero angle. This peak is significantly narrower than that for 
signal events. This happens, in particular, because the
emission of the photons by the electron from $K_{e3}$ decay
occurs in the setup material downstream the decay vertex,
but angle is still calculated as if emission comes from the vertex.

Cut4: \hspace{0.8cm} $0.002 < \theta_{e \gamma}$ $ <0.030$ 

Right part of this cut is introduced for suppression of
background channels(1,2,4,5).
After all cuts 6079 event are selected, with a background of 1603 events.
Background normalization is done by comparison numbers of events for  
$K_{e3}$ decay 
in MC and real data samples.

Event reductions statistics are summarized in Table 1.  

\begin{figure}
\includegraphics[scale=.60]{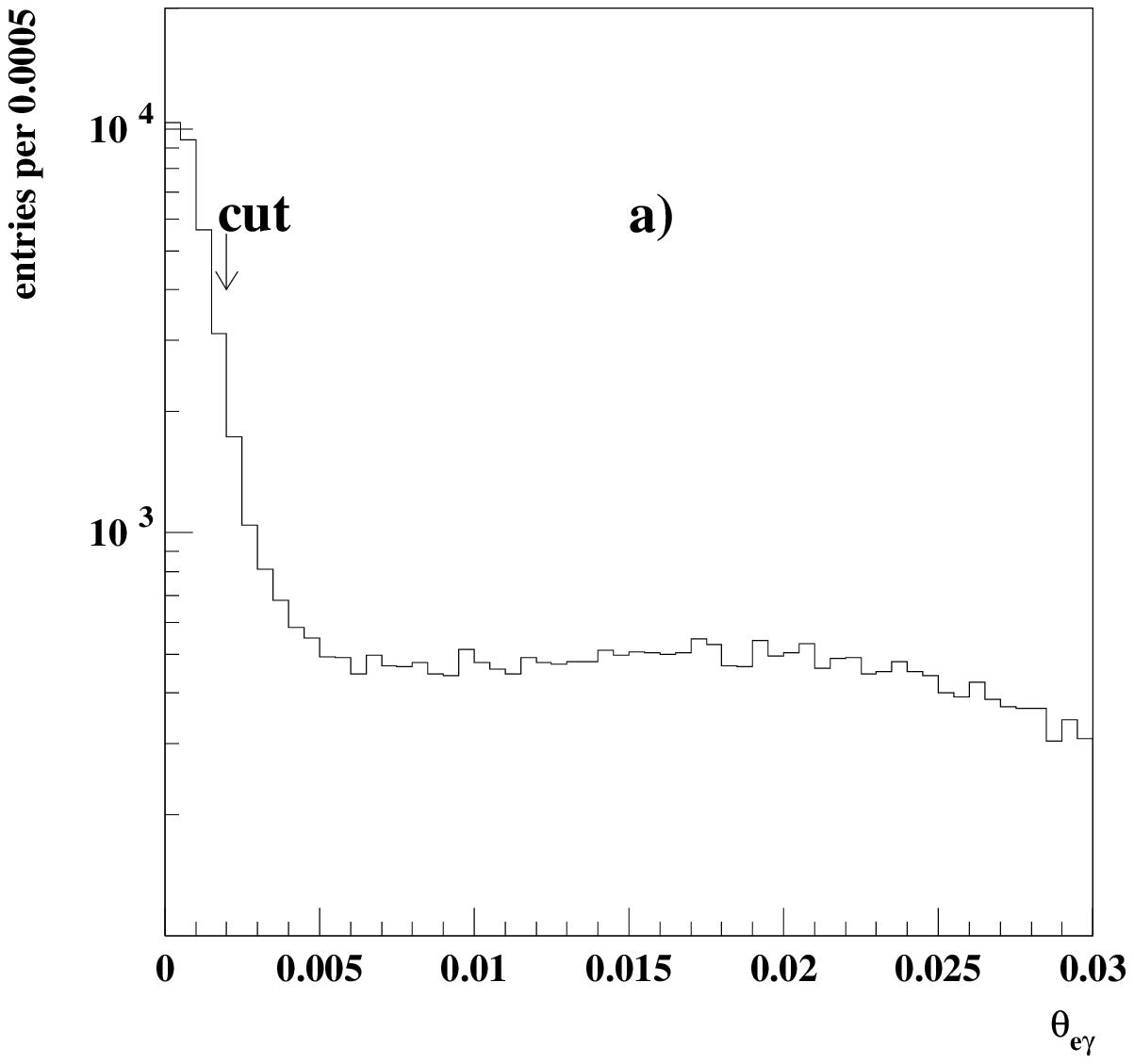}
\includegraphics[scale=.60]{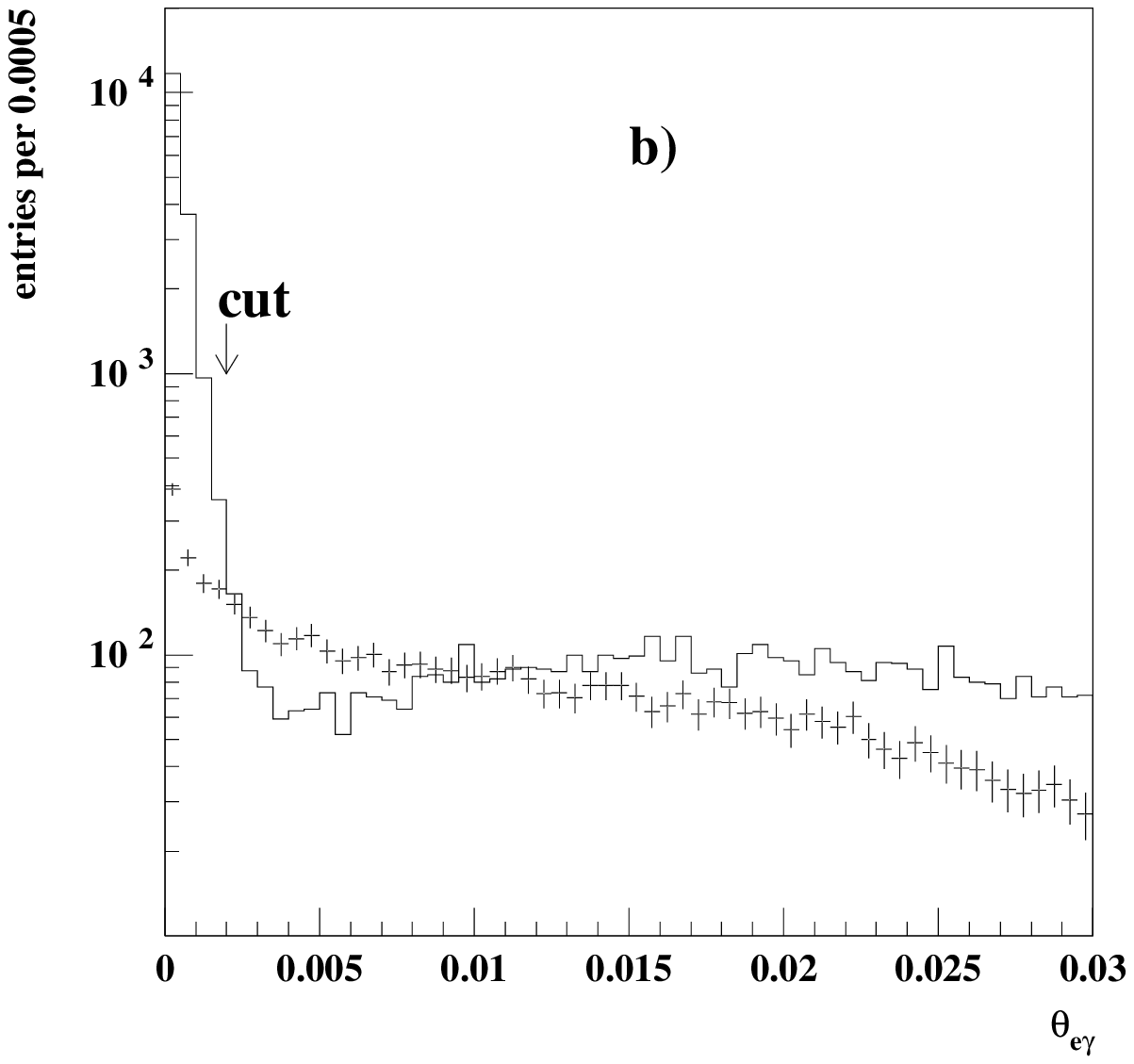}
\caption{\em Distribution    over 
$\theta_{e \gamma}$ - the angle between electron and photon in lab.
system. a) real data; b)MC background (histogram) and 
signal(points with errors)}
\label{figure:piz9}
\end{figure}

\begin{table}
\begin{center}
\begin{tabular}{|l|r|r|r|}
\hline
Cut & real data & background& signal MC   \\

\hline

Number of events selected                    &  41072  &  32901  & 11180 \\
$E_{miss} > 0.5GeV$                          &  37428  &  31134  & 10035 \\
$-0.01 < M^{2}(\pi^{0} e^{-} \gamma) < 0.01$ &  26277  &  25287  & 8430  \\
 $0.09 < M^{2}(\pi^{-}\pi^{0}) < 0.24 $      &  23293  &  21648  & 7153  \\
$0.002 < $$\theta_{e \gamma} < 0.030$        &   6079  &   1603  & 4476  \\
\hline
\end{tabular}
\end{center}
\caption{\em  Event reduction statistics for the real data, the background 
MC and signal MC.}
\label{table:avmul}
\end{table}

\section {Results }

\vspace{0.5cm}
\begin{table}
\begin{center}
\begin{tabular}{|c|r|c|}
\hline
$R_{exp}\times 10^{2}$ & ev numb & experiment   \\

\hline

$ 0.47\pm0.02 $&  1456  & this exp. \\

$ 0.46\pm0.08 $&    82  & XEBC \cite{e4}     \\
$ 0.56\pm0.04 $&   192  & ISTRA \cite{e5}     \\
$ 0.76\pm0.28 $&    13  & HLBC \cite{e6}     \\

\hline
\end{tabular}
\end{center}
\caption{\em $Br(K^- \rightarrow \pi^{0} e^{-} \overline\nu_{e} \gamma)/
            Br(K^- \rightarrow \pi^{0} e^{-} \overline\nu_{e})$ for
 $E(\gamma) > 10 MeV, 0.6<cos\theta_{e\gamma}<0.9$  in comparison with
 previous data. }
\label{table:avrrmul}
\end{table}

The resulting distribution of the selected events over 
$cos\theta{^*}_{e\gamma}$, $\theta{^*}_{e\gamma}$ being the
angle between the electron and the photon in
the kaon rest frame is shown in Fig.8. The distribution over  
$E{^*}_{e\gamma}$ - the photon energy in the kaon rest
frame is shown in Fig. 9.  Reasonable agreement of the
date with MC is seen. When generating the signal MC, a generator 
based on $O(p^2)$ \cite{v10} is used. 

\begin{figure}
\includegraphics[scale=.60]{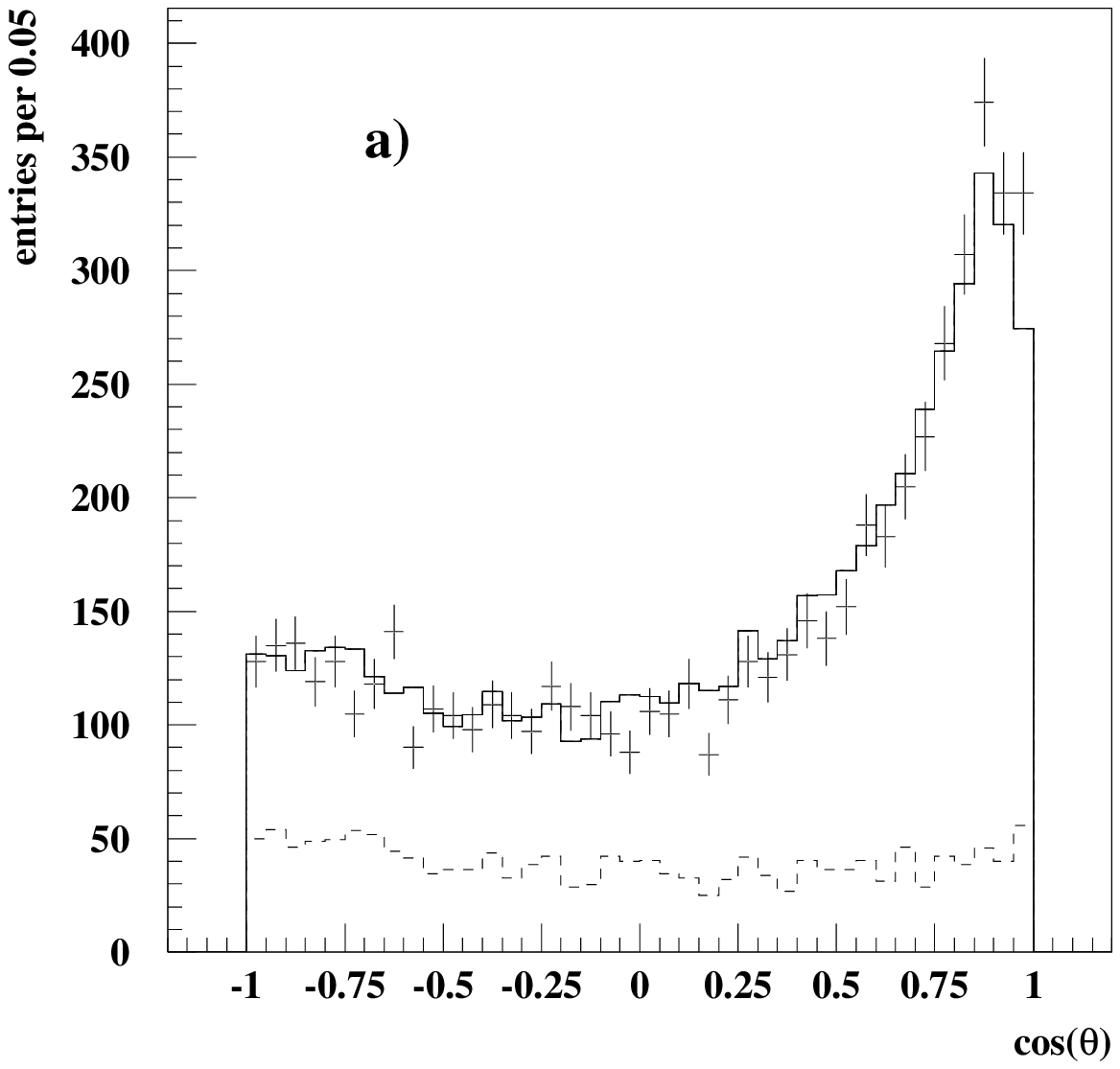}
\includegraphics[scale=.60]{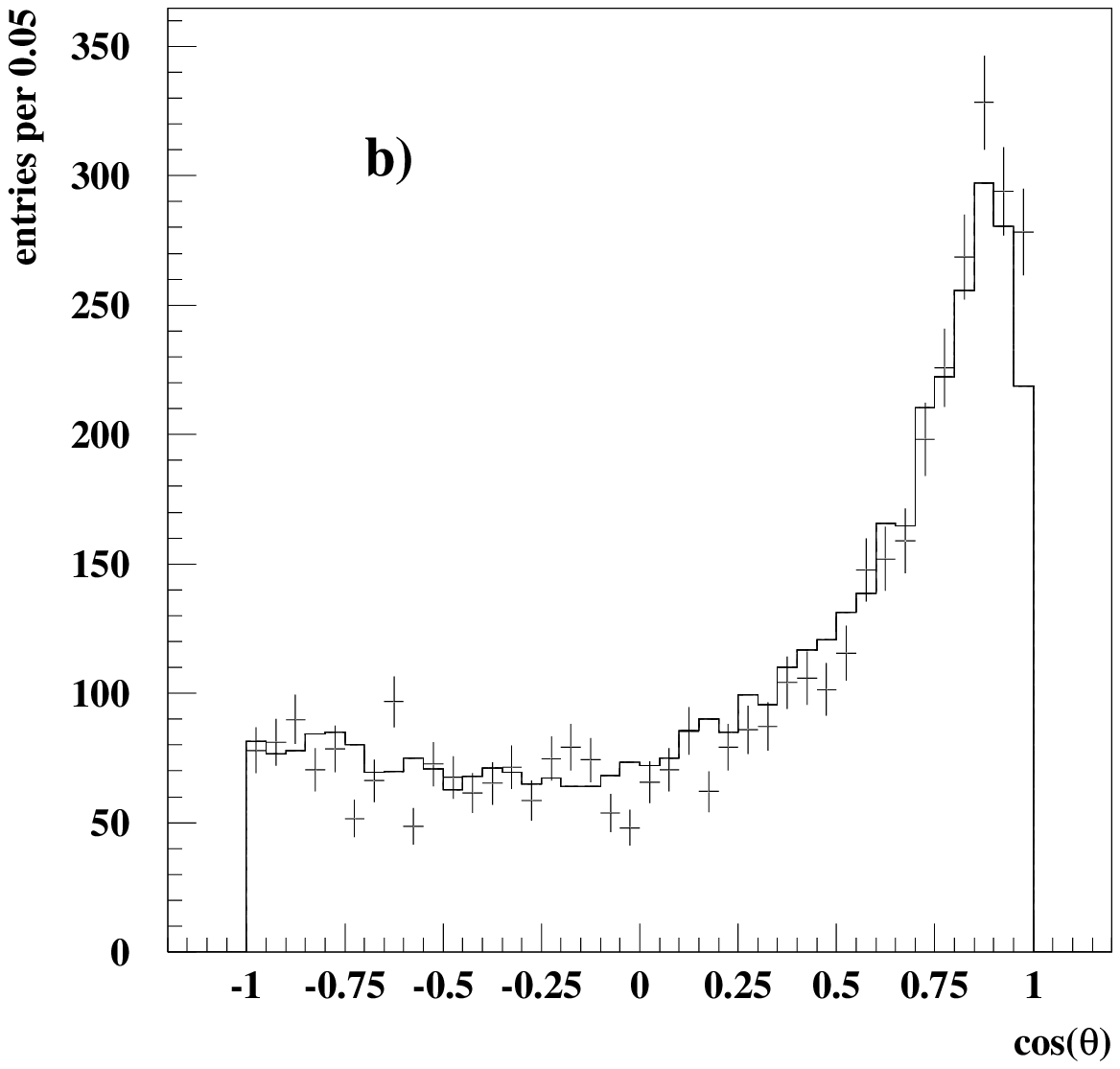}
\caption{\em a) The distribution of the events over $cos\theta^{*}_{e\gamma}$.
Points with errors are the real data, histogram is  -
total MC signal plus background.
Dotted line histogram is background.
b)the same after background subtraction}
\label{figure:piz7}
\end{figure}

\begin{figure}
\centering
\includegraphics[scale=.60]{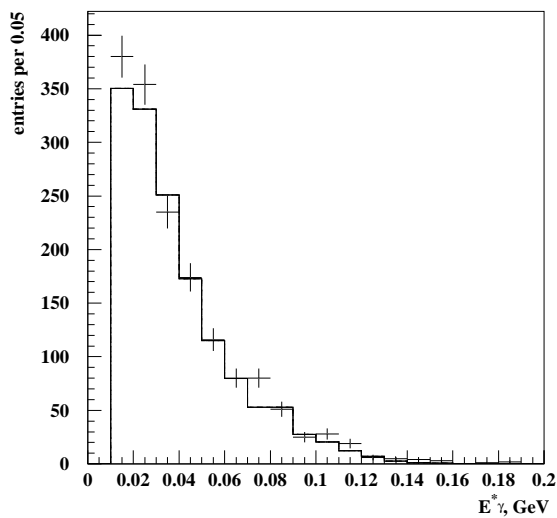}
\caption{\em The distribution of the events over $E^{*}_{\gamma}$-   
 the energy of the photon in the kaon rest frame. Histogram corresponds to the
 data, points with errors- total signal plus background MC.  }
\label{figure:piz16}
\end{figure}

To obtain the branching ratio of the $K_{\pi^{0} e^{-} 
\overline\nu_{e}\gamma}$ relative to the  $K_{e3}$ (R), 
the background and efficiency corrected number of
$K_{e3\gamma}$ events is compared to that of 569923~~ $K_{e3}$ events
found with the similar selection criteria. The branching ratio (R) is found to be
\begin{equation}
   R= \frac{Br(K^- \rightarrow \pi^{0} e^{-} \overline\nu_{e}\gamma)}
{Br(K^- \rightarrow \pi^{0} e^{-} \overline\nu_{e})}
  =(1.81\pm0.03(stat)\pm0.07(syst) )\cdot10^{-2}
\end{equation}
for $E^{*}_{\gamma}>10MeV$ and $\theta^{*}_{e\gamma} > 10^{\circ}$.
Systematic errors are estimated by variation of the cuts of Table 1.

For comparison with previous experiment the branching ratio
with cuts $E^{*}_{\gamma}>10$MeV,
$0.6<cos\theta^{*}_{e \gamma}<0.9$ is calculated
\begin{equation}
   R= \frac{Br(K^- \rightarrow \pi^{0} e^{-} \overline\nu_{e} \gamma) }
{Br(K^- \rightarrow \pi^{0} e^{-} \overline\nu_{e})}
=(0.47\pm0.02(stat)\pm0.03(syst))\cdot10^{-2}
\end{equation}

The results of previous experiments are given in  Table.2

For the cuts  $E^{*}(\gamma) > 30 MeV$ and $\theta^{*}_{e\gamma} > 
20^{\circ}$,
used in most theoretical papers 

\begin{equation}
   R= \frac{Br(K^- \rightarrow \pi^{0} e^{-} \overline\nu_{e} 
\gamma)}
{Br(K^- \rightarrow \pi^{0} e^{-} \overline\nu_{e})}
  =(0.64\pm0.02(stat)\pm0.03(syst))\cdot10^{-2}.
\end{equation}
Using PDG value for  $K_{e3}$ decay branching for $K^- \rightarrow 
\pi^{0} e^{-} \overline\nu_{e} \gamma$ is calculated 
$ Br = (3.06\pm0.09\pm0.14)\cdot 10^{-4}$. 
It can be compared with theoretical prediction\cite{v3}
at tree level $ Br = 2.8 \cdot 10^{-4}$ and $ Br = 3.0 \cdot 10^{-4}$
for $O(p^4)$ level.  Theoretical prediction of 
V.V.Braguta, A.A.Likhoded, A.E.Chalov\cite{v10} at tree level  
is  $ Br = 3.12 \cdot 10^{-4}$.

For the asymmetry $A_{\xi}$(for the same cuts as in Table.2) we get
\begin{equation}
A_{\xi} = -0.015 \pm 0.021
\end{equation}

At present it is the best estimate of this asymmetry.
It can be compared with an upper limit on the $A_{\xi}$ value
$|A_{\xi}(K^-\rightarrow\pi^{0} e^{-} \overline \nu_{e} \gamma)|< 0.8\cdot10^{-4}$
in the $SU(2)_{L} \times SU(2)_{R} \times U(1)$ model\cite{v14}
and $ A_{\xi} = - 0.59 \cdot 10^{-4}$ in the Standard Model\cite{v10}

\vspace{1cm}

The authors would like to thank D.S. Gorbunov,
V.A. Matveev and V.A. Rubakov, 
for numerous discussions.   
V.V.Braguta, A.A.Likhoded, A.E.Chalov for program of matrix
element calculation.
The work is supported in part by the RFBR
grants N03-02-16330 (IHEP group) and N03-02-16135 (INR group)
and by Russian Science Support Foundation (INR group).


\end{document}